\documentclass[10pt,cspaper,compsoc]{IEEEtran}
\ifCLASSOPTIONcompsoc
  \usepackage{cite}
\else
  \usepackage{cite}
\fi
\ifCLASSINFOpdf
  \usepackage[pdftex]{graphicx}
  \graphicspath{{./trees/}}
  \DeclareGraphicsExtensions{.pdf}
\else
\fi
\usepackage[cmex10]{amsmath}
\ifCLASSOPTIONcompsoc
  \usepackage[caption=false,font=normalsize,labelfont=sf,textfont=sf]{subfig}
\else
  \usepackage[caption=false,font=footnotesize]{subfig}
\fi
\usepackage{booktabs}
\usepackage{rotating}

\usepackage{amsthm}

\usepackage{amsfonts}

\usepackage{pstricks,pst-node,pst-tree}
\usepackage{amsmath}
\usepackage{amssymb}
\usepackage{xcolor}

\usepackage{pstricks}

\newtheorem{lemm}{Lemma}
\newtheorem{prop}{Proposition}

\definecolor{myblue}{rgb}{0,0.0,0.0}
\definecolor{myred}{rgb}{0.9,0.1,0}
\newcommand{\fil}[1]{\textcolor{myblue}{#1}}

\begin{document}
%
\title{On the number of ranked species trees \\ producing anomalous ranked gene trees}
%
%
%
%

\author{Filippo~Disanto %
\thanks{F.~Disanto is with the Department of Biology, Stanford University, Stanford, CA, USA. Email: fdisanto@stanford.edu.} and
Noah A.~Rosenberg %
\thanks{N.~A.~Rosenberg is with the Department of Biology, Stanford University, Stanford, CA, USA. Email: noahr@stanford.edu.} }

\IEEEcompsoctitleabstractindextext{%
\begin{abstract}
Analysis of probability distributions conditional on species trees has demonstrated the existence of anomalous ranked gene trees (ARGTs), ranked gene trees that are more probable than the ranked gene tree that accords with the ranked species tree. Here, to improve the characterization of ARGTs, we study enumerative and probabilistic properties of two classes of ranked labeled species trees, focusing on the presence or avoidance of certain subtree patterns associated with the production of ARGTs. We provide exact enumerations and asymptotic estimates for cardinalities of these sets of trees, showing that as the number of species increases without bound, the fraction of all ranked labeled species trees that are ARGT-producing approaches $1$. This result extends beyond earlier existence results to provide a probabilistic claim about the frequency of ARGTs.
\end{abstract}

}

\maketitle

\IEEEdisplaynotcompsoctitleabstractindextext

%
\IEEEpeerreviewmaketitle

\section{Introduction}

\fil{Recent research in phylogenetics has conducted detailed probabilistic explorations of the properties of different gene tree structures using models of gene lineage evolution conditional on species trees \cite{AllmanDegnanRhodes11:jtb, AllmanDegnanRhodes11:jmb, Degnan13, DegnanAndSalter05, DegnanEtAl09}. These phylogenetic modeling investigations uncover new phylogenetic phenomena, facilitate mathematical and simulation-based analyses of complex data spaces for phylogenetic studies, enable development and theoretical analysis of species tree inference algorithms, and assist in identifying strengths, limitations, and protocols for proposed methods \cite{DegnanAndRosenberg09, PamiloAndNei88, Rosenberg13:mbe, RosenbergAndTao08, Wu12}.}

A ranked labeled gene tree, or gene tree labeled history, consists of a \fil{rooted} labeled gene tree topology together with the temporally ordered sequence in which coalescences in the gene tree take place \cite{Harding71, Song06}. \fil{Ranked gene trees arise in a model of random bifurcation in which each lineage is equally likely to be the next to bifurcate, or, backward in time, each pair of lineages is equally likely to be the next to coalesce. This simple branching assumption, originating from the classical \emph{Yule model}~\cite{Yule24} and providing the model of tree topology in coalescent models for gene lineage evolution \cite{HeinEtAl05, Wakeley09}, generates a convenient uniform distribution on the set of ranked gene trees \cite{Edwards70, Page91}.}

\fil{Given a genealogical history of a set of gene lineages, the ranked gene tree is an elemental tree structure, in the sense that other structures---such as unranked rooted gene trees, unranked unrooted gene trees, and the list of clades included in a tree---are uniquely specified by a ranked gene tree, whereas many ranked gene trees might be compatible with a given choice for one of these other structures. Thus, as properties of other structures can often be derived from properties of ranked gene trees \cite{Brown94, DisantoEtAl13, SteelAndMcKenzie01, Rosenberg06:anncomb}, ranked gene trees represent a natural class of objects for phylogenetic modeling.}

\fil{Degnan {\it et al.}~\cite{DegnanEtAl12:mathbiosci} initiated the probabilistic study of ranked gene trees in species tree models, providing a formula under the standard multispecies coalescent model \cite{DegnanAndRosenberg09, DegnanAndSalter05, Maddison97, PamiloAndNei88, Rosenberg02} for the probability conditional on a labeled species tree that a particular ranked gene tree is produced (see also \cite{StadlerAndDegnan12}). Under the model, \cite{DegnanEtAl12:mathbiosci} termed} ranked labeled gene trees that are more likely to be generated than the ranked labeled gene tree that matches the ranked labeled species tree \emph{anomalous ranked gene trees} (ARGTs). ARGTs represent a surprising outcome of genealogical descent in which an unexpected ranked gene tree exceeds the model ranked species tree in probability.

Degnan {\it et al.}~\cite{DegnanEtAl12:tcbb} obtained a full characterization of the set of unranked labeled species trees for which at least one ranking produces ARGTs.  That is, they identified all unranked labeled species trees for which a ranking and a set of branch lengths can be selected so that the most likely ranked gene tree conditional on the ranked species tree together with its branch lengths disagrees with the ranked species tree. They found that the set of unranked labeled species trees with at least one ARGT-producing ranking is precisely the set of unranked labeled species trees that do not have a caterpillar or pseudocaterpillar shape.

While the constructive proof of \cite{DegnanEtAl12:tcbb} identifies specific ARGT-producing rankings for a given unranked labeled species tree, the set of \emph{ranked} labeled species trees that are ARGT-producing remains incompletely characterized. For small trees, Table 1 of \cite{DegnanEtAl12:tcbb} reported the numbers of ranked labeled species trees that give rise to ARGTs, but general results have not been presented to assess the fraction of ranked labeled species trees that are ARGT-producing.

Here, we show that as the number of species increases without bound, the fraction of all ranked labeled species trees that are ARGT-producing\fil{---that is, the fraction for which some set of species tree branch lengths gives rise to ARGTs---}approaches $1$.  In other words, we extend beyond the proof of \cite{DegnanEtAl12:tcbb} to argue that not only does each unranked species tree have at least one ARGT-producing ranking, nearly all ranked species trees are ARGT-producing. We obtain the result through a combinatorial approach, counting the number of ranked labeled species trees with $n$ internal nodes that are identified by the proof of \cite{DegnanEtAl12:tcbb} as ARGT-producing, and we show that the ratio of the cardinality of this set and the total number of ranked labeled \fil{species} trees on $n$ nodes, or $(n+1)! \, n!/2^n$, approaches 1 as $n$ approaches infinity.

\section{Preliminaries}\label{prel}

\noindent \textbf{Ranked trees, ranked species trees, and ordered ranked trees.} It is convenient here to index tree and subtree sizes by the number of internal nodes, rather than by the usual index, the number of leaves.

A \emph{ranked} tree $t$ of size $n$ is a binary rooted tree with $n$ \emph{internal} nodes (and $n+1$ leaves), each one bijectively associated with a number in $\{1, 2, ..., n  \}$.  The labeling of the internal nodes must be \emph{increasing}, in the sense that each path from the root of $t$ to a leaf contains an increasing sequence of numbers. The increasing labeling gives a time ordering of the coalescence events occurring along the branches of the tree. The most recent event is the one that carries the greatest label. Ranked trees are considered in a graph-theoretic sense. Therefore, unless specified otherwise, they do not carry any left-right orientation.

A ranked \emph{species} tree is a ranked tree equipped with a labeling for its taxa. Thus, two ranked species trees can be the same when treated as ranked trees but different in their leaf labeling. The set of ranked species trees is denoted by $\mathcal{S}$, and $\mathcal{S}_n$ denotes the set of ranked species trees of size $n$. It is well-known (\cite{Rosenberg06:anncomb}, Corollary~3.2) that the cardinality of $\mathcal{S}_n$ is
\begin{equation}\label{spectr}
|\mathcal{S}_n| = \frac{(n+1)! \, n!}{2^n}.
\end{equation}

\begin{figure}[tb]
\begin{center}
\includegraphics[width=\columnwidth]{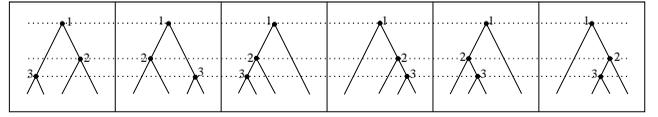}
\end{center}
\caption{{\small The six ordered ranked trees of size $n=3$ internal nodes. Left-right orientation determines different trees. }}\label{sei}
\end{figure}

An \emph{ordered} ranked tree is a ranked tree provided with a left-right orientation of its subtrees. The set of ordered ranked trees is denoted by $\mathcal{R}$, and $\mathcal{R}_n$ is the subset of $\mathcal{R}$ consisting of those trees of size $n$. The cardinality of $\mathcal{R}_n$ is (\cite{FlajoletAndSedgwick09}, Example~II.17) 
\begin{equation}\label{ordtr}
|\mathcal{R}_n| = n!.
\end{equation}
In Fig.~\ref{sei}, we depict the six ordered ranked trees of size $3$. Note that in each tree, the labeling of the internal nodes increases from the root toward the leaves.

\medskip

\noindent \textbf{Maximally probable and non-maximally probable subtrees.}
Following Proposition~6 of \cite{DegnanEtAl12:tcbb}, given a ranked tree $t$ and an internal node $k$, we say that $k$ generates a \emph{maximally probable} subtree (MP-subtree for short) if we can assign the name $L$ to one of the two subtrees appended to node $k$ and the name $R$ to the other such subtree in such a way both (i) and (ii) hold for that assignment:

\begin{enumerate}
\item[(i)] $m \geq q \geq 0$, where $m = |L|$ and $q = |R|$.
\item[(ii)] Looking back in time, the sequence of coalescences in the subtree of node $k$ has the form
\begin{equation}\label{potta}
\ell^{m-q}\{\ell r, r \ell\}^q,
\end{equation}
where $\ell$ and $r$ stand for coalescence events belonging to subtrees $L$ and $R$, respectively.
\end{enumerate}

\noindent The notation $\{a,b \}^q$ in (\ref{potta}) indicates the set of words of length $q$ over the alphabet $\{a,b\}$, where $a=\ell r$ and $b = r \ell$. Thus, by $\ell^{m-q}\{\ell r, r \ell\}^q$, for $m \geq q$, it is meant that the first $m-q$ entries are in $L$, after which $q$ pairs of entries appear. Each pair has one event in $L$ and the other in $R$, and the sequences of these events within pairs are not necessarily the same.  The suggestive labels $L$ and $R$ can refer to the \emph{left} and \emph{right} subtrees of $k$, but the definition of maximally probable does not require specification of which subtree is denoted $L$ and which is denoted $R$.

Given a ranked tree $t$ and an internal node $k$, we say that $k$ generates a \emph{non-maximally probable} subtree (NMP-subtree for short) when it does not generate an MP-subtree. It is equivalent for a ranked species tree $t$ to avoid NMP-subtrees and to contain only MP-subtrees.  The subset of trees in $\mathcal{S}$ \fil{containing only MP-subtrees is denoted $\mathcal{S}^{(mp)}$}. By $\mathcal{S}_n^{(mp)}$, we indicate trees in $\mathcal{S}^{(mp)}$ of size $n$.

\begin{figure}
\begin{center}
\includegraphics*[width=.6\columnwidth]{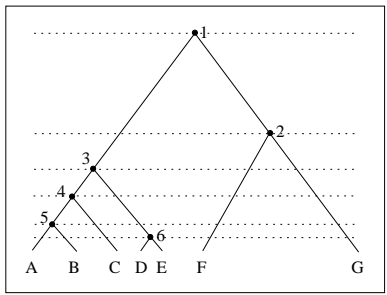}
\end{center}
\caption{{\small A ranked species tree that is non-maximally probable (NMP) at internal node $3$. This tree is maximally probable (MP) at the root.}}\label{nmp}
\end{figure}

The tree in Fig.~\ref{nmp} contains exactly one NMP-subtree, that is, the one generated at node $3$. Indeed, observe that the only possible assignment of $L$ and $R$ that satisfies (i) gives a sequence of coalescences $r \ell \ell$ that does not match (\ref{potta}); none of the other nodes generates an NMP-subtree. For instance, at node $1$, we can assign $L$ to the subtree generated by node $3$ and $R$ to the subtree generated by node $2$, and the resulting sequence of coalescence events is $\ell \ell \ell \ell r$.

\begin{figure}[tpb]
\begin{center}
\includegraphics*[width=.72\columnwidth]{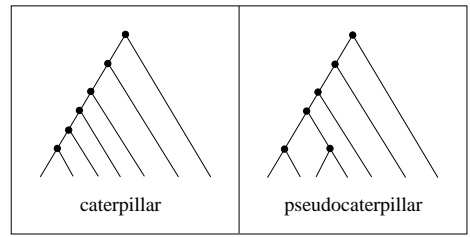}
\end{center}
\caption{{\small Caterpillar and pseudocaterpillar trees. These trees do not contain NMP-subtrees.}}\label{pseudo}
\end{figure}

Note that for a node $k$ to generate an NMP-subtree it is necessary to satisfy the following \emph{1-2 condition}: one of the two subtrees appended to $k$ has size at least $1$ and the other has size at least $2$. Trees for which the 1-2 condition is not satisfied for \fil{any} internal node are either \emph{caterpillar} or \emph{pseudocaterpillar} (Fig.~\ref{pseudo}), using the definition that a tree has a \emph{caterpillar} shape when each internal node has at least one leaf stemming from it, and a \emph{pseudocaterpillar} shape when it is not a caterpillar and, still, no node has the 1-2 condition.

We define $\mathcal{S}^{(cat)}$ as the set of caterpillar and pseudocaterpillar ranked species trees. The subset $\mathcal{S}_n^{(cat)}$ contains such trees of size $n$. Caterpillar and pseudocaterpillar trees are not NMP, and they contain no NMP-subtrees.

\medskip

\noindent \textbf{Anomalous ranked gene trees.} We recall that an \emph{anomalous} ranked gene tree (ARGT) is a ranked gene tree that does not match the ranked species tree and that has probability under the multispecies coalescent model greater than that of the matching ranked gene tree \cite{DegnanEtAl12:tcbb, DegnanEtAl12:mathbiosci}.  We say that a ranked species tree produces ARGTs if there exist values for the speciation times such that the ranked species tree together with the speciation times has at least one ARGT.

When we disregard the ranking of the coalescences in the species tree, the set of unranked species trees that produce ARGTs has a known complete characterization. In particular, as shown in Theorem~1 of \cite{DegnanEtAl12:tcbb}, each unranked species tree $t$ that is neither a caterpillar nor a pseudocaterpillar can be ranked in such a way that it is NMP at a particular subtree $H(t)$. Further, being NMP at a subtree implies that speciation times can be chosen to produce an ARGT at that subtree. Thus, each unranked species tree $t$ other than caterpillars and pseudocaterpillars produces ARGTs.

Here, we focus on \emph{ranked} species trees that produce ARGTs. That is, the ranking of the species tree is given and it cannot be carefully selected as in the unranked case studied by \cite{DegnanEtAl12:tcbb} and \cite{DegnanEtAl12:mathbiosci}. Formally, from Propositions~9, 2, and 3 of \cite{DegnanEtAl12:tcbb}, we borrow two facts:
\begin{enumerate}
\item[(iii)] If a ranked species tree $t$ contains an NMP-subtree, that is, $t \in \mathcal{S} \setminus \mathcal{S}^{(mp)}$, then $t$ produces ARGTs at the NMP-subtree.
\item[(iv)] If a ranked species tree $t$ is either a caterpillar or a pseudocaterpillar, that is, $t \in \mathcal{S}^{(cat)}$, then $t$ does not produce ARGTs.
\end{enumerate}

\noindent As stated in \cite{DegnanEtAl12:tcbb}, (iii) is only a sufficient condition for production of ARGTs and not a complete characterization of the set of ranked species trees that generate ARGTs. Because (iii) connects NMP-subtrees to the problem of counting ranked species trees that produce ARGTs, our interest is in counting ranked species trees containing or avoiding NMP-subtrees.
\medskip

\noindent \textbf{A subtree specified by the 1-2 condition.} Property (iii) states that being NMP at a given subtree implies producing ARGTs at that particular subtree. It is of interest to investigate not only the presence of ARGT-producing subtrees but also their position in the \fil{species} tree.  Here we introduce the set of ranked species trees $t$ for which (iii) ensures production of ARGTs at the largest subtree $H(t)$ that satisfies the 1-2 condition. In particular, for any ranked species tree $t$, there is no NMP-subtree that properly contains $H(t)$. It is by examining the ranking of $H(t)$ that \cite{DegnanEtAl12:tcbb} showed that with the exception of caterpillars and pseudocaterpillars, each unranked species tree produces ARGTs.

The subtree $H(t)$ can be defined by a recursive query procedure: starting from the root of the tree $t$, if the current node satisfies the 1-2 condition, then stop and set $H(t)$ equal to the subtree rooted at the current node. Otherwise, at the current node, the tree splits into two subtrees that either both have size smaller than $2$, or exactly one of them has size smaller than $1$. In the first case, stop the procedure and set $H(t)$ empty. In the second case, query the node whose subtree has at least size $2$. Observe that $H(t)$ is empty if and only if $t$ is either a caterpillar or a pseudocaterpillar. The symbol $\mathcal{S}^{(H)}$ denotes the set of ranked species trees $t$ that are MP at $H(t)$. The tree in Fig.~\ref{nmp} belongs to $\mathcal{S}^{(H)}$ but not to $\mathcal{S}^{(mp)}$; the subtree $H(t)$ is, in this case, the subtree generated by the root.

As was observed in \cite{DegnanEtAl12:tcbb},
\begin{equation}\label{gerarchia}
\mathcal{S}^{(cat)} \subseteq \mathcal{S}^{(mp)} \subseteq \mathcal{S}^{(H)}.
\end{equation}
Thus, $|\mathcal{S}_n| - |\mathcal{S}_n^{(H)}|$ bounds from below the cardinality of $\mathcal{S}_n \setminus \mathcal{S}_n^{(mp)}$, also providing a lower bound for the ultimate quantity of interest, the number of ranked species trees that produce ARGTs.

\section{Results}\label{ultima}

We now present enumerative results for the classes of ranked species trees that we have introduced. In Section~\ref{vauno}, we show that the probability that a randomly selected ranked species tree of size $n$ produces ARGTs approaches $1$ as $n$ becomes large. In Section~\ref{acca}, we obtain the enumeration of the set $\mathcal{S}_n^{(H)}$. Section~\ref{double} provides a recursion to enumerate $\mathcal{S}_n^{(mp)}$. The recursion enables a closed formula that bounds from below the number of ARGT-producing ranked species trees of size $n$. First, in Section~\ref{equivalenceLemma}, we obtain a result that allows us to switch our perspective between ranked species trees and ordered ranked trees.

\subsection{Equivalence between ranked species trees and ordered ranked trees} \label{equivalenceLemma}

Observe that the subtree patterns defining $S^{(cat)}$, $S^{(mp)}$, and $S^{(H)}$ do not depend on the leaf labeling, and only consider the ranking of the internal nodes. To simplify our computations, we focus on ordered ranked trees instead of ranked species trees, using an equivalence to convert results about ordered ranked trees into results about ranked species trees. If $P$ is a tree property that does not concern labeling of taxa but only concerns the ranking of the coalescence events---such as avoiding NMP-subtrees, for instance---then the two sets of trees can be treated as equivalent. More precisely, we have the following:
\begin{prop}\label{equ}
If $P$ is a tree property that depends only on the ranking of the coalescence events, then
\begin{equation}\label{poppa}
\frac{ | \{t \in \mathcal{R}_n : P(t) \} |   }{n!}=\frac{  | \{ t \in \mathcal{S}_n : P(t)  \} |  }{(n+1)!\,n!/2^n}.
\end{equation}
\end{prop}
\emph{Proof.} Define an equivalence relation $\approx_o$ on the set of ordered ranked trees of the same size, so that $t_a \approx_o t_b$ when $t_b$ can be obtained from $t_a$ by switching pairs of subtrees appended to corresponding nodes---in other words, if, ignoring left-right orientation, $t_a$ and $t_b$ represent the same ranked tree. Similarly, define the equivalence relation $\approx_s$ on the set of ranked species trees of the same size, so that $t_c \approx_s t_d$ if $t_c$ and $t_d$ represent the same ranked tree once labels for the leaves have been removed.

On the set of ranked trees of size $n$, consider the probability distribution induced by the \fil{Yule model of random branching}. Under this model, the probability of a ranked tree $t$ depends on two parameters: the size $n$ and the number of subtrees of size $1$ (i.e.~cherries), denoted by $c(t)$. We have $P_{\text{Yule}}(t) = 2^{n-c(t)}/n!$, as in Theorem~3.4 of \cite{Rosenberg06:anncomb} (see also \cite{Page91, Tajima83}).

Observe that for a fixed ordered ranked tree $t$ of size $n$, the cardinality of the equivalence class $[t]_{\approx_o}$ is given by $2^{n-c(t)}$ because switching left and right subtrees at the root of a subtree of size greater than $1$ is the only way to produce a different ordered ranked tree. Similarly, if we fix a ranked species tree $t$, then the cardinality of $[t]_{\approx_s}$ is $(n+1)!/2^{c(t)}$. Indeed, each of the possible $(n+1)!$ permutations of the leaf labels of $t$ gives exactly $2^{c(t)}$ equivalent labelings of the taxa.

It follows that if we fix the size $n$, then the uniform distribution over the set of ordered ranked trees and the uniform distribution over the set of ranked species trees induce the same probability distribution---the Yule distribution---over the set of ranked trees. In particular, the probability of a ranked tree under the Yule model is given by the cardinality of the corresponding equivalence class in $\approx_o$ divided by $n!$, or by the cardinality of the equivalence class in $\approx_s$ divided by  $(n+1)! \, n!/2^n$.

Finally, observe that the property $P$ respects the equivalence classes defined under $\approx_o$ and $\approx_s$ in the sense that an ordered ranked tree (resp.~ranked species tree) $t$ satisfies $P$ if and only if all the ordered ranked trees (resp.~ranked species trees) in the equivalence class $[t]_{\approx_o}$ (resp.~$[t]_{\approx_s}$) satisfy $P$.

We can then write
\begin{eqnarray}\nonumber
\frac{ | \{t \in \mathcal{R}_n : P(t) \} |   }{n!} &=& \sum_{[t]_{\approx_o} \text{\,:} P(t)} \frac{|[t]_{\approx_o}|}{n!} \\\nonumber
&=& \sum_{[t]_{\approx_s} \text{\,:} P(t)} \frac{|[t]_{\approx_s}|}{(n+1)!\,n!/2^n} \\\nonumber
&=& \frac{  | \{ t \in \mathcal{S}_n : P(t)  \} |  }{(n+1)!\,n!/2^n}. \,\,\, \Box \nonumber
\end{eqnarray}

\medskip

In the framework of ordered trees, we define $\mathcal{R}^{(mp)}$, $\mathcal{R}^{(H)}$, and $\mathcal{R}^{(cat)}$ as corresponding versions of the classes $\mathcal{S}^{(mp)}$, $\mathcal{S}^{(H)}$, and $\mathcal{S}^{(cat)}$, respectively. Indeed, our definitions for sets $\mathcal{S}^{(x)}$ did not depend on the left-right orientation of subtrees. Therefore, the same definitions apply to ordered ranked trees to define the associated $\mathcal{R}^{(x)}$. To determine the cardinality of a set $\mathcal{S}_n^{(x)} \subseteq \mathcal{S}_n$, our approach consists of finding the cardinality of the corresponding ordered set $\mathcal{R}_n^{(x)} \subseteq \mathcal{R}_n$ and then applying (\ref{poppa}) to obtain
\begin{equation}\label{collina}
|\mathcal{S}_n^{(x)}| = \frac{(n+1)!}{2^n} \, |\mathcal{R}_n^{(x)}|.
\end{equation}

\subsection{Probability that a ranked species tree produces ARGTs}\label{vauno}
We are now ready to show that the probability that a randomly selected ranked species tree of size $n$ produces ARGTs approaches $1$ as $n$ becomes large. It is useful to introduce the sequence $\alpha_n$, defined as
\begin{equation}\label{alfia}
\alpha_n = \sum_{q=1}^{n-1}  \frac{2^{\min(q,n-q)}}{{{n}\choose{q}}}.
\end{equation}
Considering $q=1$ and $q=n-1$ in the sum, we find
\begin{equation}\label{pinco}
\alpha_n \geq 4/n.
\end{equation}
We also have
$$\alpha_n \leq 2 \sum_{q=1}^{\lfloor n/2 \rfloor} \frac{2^q}{{{n}\choose{q}}} \leq 2 \sum_{q=1}^{\lfloor n/2 \rfloor} \frac{2^q}{{{2 \lfloor n/2 \rfloor }\choose{q}}} = 2 s_{\lfloor n/2 \rfloor}.$$
The sequence $$s_n = \sum_{q=1}^{n} \frac{2^q}{{{2 n}\choose{q}}}$$ can be bounded by
\begin{equation}\label{pincio}
s_n \geq 1/n,
\end{equation}
considering only the $q=1$ term in the sum. Furthermore, $s_n$ has the following property.
\begin{lemm}
The sequence $s_n$ satisfies the recursion
\begin{equation}\label{recu}
9(2n+1)s_{n+1}-4(2n+3)s_n = \frac{10 n + 9}{n+1} + \frac{n (2^{n+1})}{{{2n}\choose{n}}}.
\end{equation}
\end{lemm}
\emph{Proof.} Using the Wilf-Zeilberger summation approach \cite{PetkovsekEtAl96}, define $F(q,n) = 2^q/{{2n}\choose{q}}$ and $$R(q,n) = \frac{(2n+1-q)[3q(2n+1) - 2(2n+1)(5n+6)]}{2(n+1)(2n+1)}.$$
It is easily verified that
\begin{eqnarray}\label{piolloi}
&& 9(2n+1)F(q,n+1) - 4(2n+3)F(q,n) \nonumber \\
&& = F(q+1,n)R(q+1,n) - F(q,n)R(q,n). 
\end{eqnarray}
Indeed, the identity follows by noting the ratios $$\frac{F(q,n+1)}{F(q,n)} = \frac{(2n+2-q)(2n+1-q)}{2(n+1)(2n+1)}$$ and $$\frac{F(q+1,n)}{F(q,n)} = \frac{2(q+1)}{2n-q}.$$

Summing both sides of (\ref{piolloi}) from $q=1$  to $n+1$, the right-hand side telescopes, giving a final contribution of $F(n+2,n)R(n+2,n) - F(1,n)R(1,n)$. Therefore,
\begin{eqnarray} \nonumber
&& 9(2n+1)s_{n+1} - 4(2n+3)\left[ s_n + \frac{2^{n+1}}{{{2n}\choose{n+1}}}  \right] \\\nonumber
&& =  \frac{10n+9}{n+1} \\\nonumber
&& \quad  - \frac{2^{2+n}(2n+1)(7n+6)(n-1)! \, (n+2)!}{(2n+2)!}, \nonumber
\end{eqnarray}
from which simple calculations lead to (\ref{recu}). \,\,\, $\Box$

\medskip

Starting from (\ref{recu}), it can be shown by induction on $n$ that for $n$ large,
\begin{equation}\label{coraggio}
s_n \leq \frac{n+10}{n(n-1)}.
\end{equation}
Consider $n \geq 23$. We can easily verify (\ref{coraggio}) for $n = 23$. For the inductive step,
we begin from a binomial inequality, which holds for $n\geq 1$ \cite{Bullen98}:
\begin{equation}\label{sturlo}
{{2n}\choose{n}} \geq \frac{2^{2n-1}}{\sqrt{n}}.
\end{equation}
We then have
$$\frac{n(2^{n+1})}{{{2n}\choose{n}}} \leq \frac{4n^{3/2}}{2^n} \leq \frac{n^4}{2^n} \leq \frac{1}{n},$$ where the last inequality holds because $n \geq 23$.
We can thus write
\begin{eqnarray}\nonumber
&& 9(2n+1)s_{n+1} - 4(2n+3)\left[\frac{n+10}{n(n-1)}\right]\\\nonumber
&& \leq 9(2n+1)s_{n+1} - 4(2n+3)s_n \\\nonumber
&& \leq \frac{10n+9}{n+1} + \frac{1}{n}, \nonumber
\end{eqnarray}
from which
$$ s_{n+1} \leq \frac{18n^3 + 100n^2 + 203n +119}{9n(n-1)(n+1)(2n+1)}.$$
Finally note that
\begin{eqnarray}\nonumber
&& \frac{(n+1)+10}{(n+1)n}  - s_{n+1} \\\nonumber
&&  \geq \frac{(n+1)+10}{(n+1)n}  - \frac{18n^3 + 100n^2 + 203n +119}{9n(n-1)(n+1)(2n+1)} \\ \nonumber
&& = \frac{89n^2 - 311n -218 }{9n(n-1)(n+1)(2n+1)}.\nonumber
\end{eqnarray}
This last quantity is positive for $n \geq 5$, completing the inductive proof of (\ref{coraggio}).

Therefore, from (\ref{pincio}) and (\ref{coraggio}), we have for $n \geq 23$,
\begin{equation}\label{pennarello}
\frac{1}{n} \leq s_n \leq \frac{n+10}{n(n-1)},
\end{equation}
producing, for $n$ large, the asymptotic equivalence
$$s_n \sim 1/n.$$

Thus, for $n$ large, by (\ref{pinco}),
\begin{equation}\label{pennarello1}
\frac{4}{n} \leq \alpha_n \leq  2 s_{\lfloor n/2 \rfloor} \sim \frac{2}{\lfloor n/2 \rfloor},
\end{equation}
so that
\begin{equation}\label{alf}
\alpha_n \sim 4/n.
\end{equation}
Table~\ref{pio} illustrates this asymptotic equivalence of $\alpha_n$ and $4/n$ for a variety of values of~$n$. It is from this asymptotic equivalence in (\ref{alf}) that the main result of this section follows.


\begin{table}
\caption{{\small Asymptotic equivalence of $\alpha_n$ and $4/n$, with $\alpha_n$ computed from (\ref{alfia}).}}
\fontsize{9}{11}\selectfont
\label{pio}
\vspace{-.2cm}
\begin{center}
\begin{tabular}{|c| c c c c c | }\hline
           & \multicolumn{5}{c|}{$n$} \\ \cline{2-6}
           & 50      &  100    & 250     & 500     & 1000      \\ \hline
$\alpha_n$ & 0.08753 & 0.04172 & 0.01626 & 0.00806 & 0.00402   \\
$4/n$      & 0.08000 & 0.04000 & 0.01600 & 0.00800 & 0.00400   \\ \hline
\end{tabular}
\end{center}
\end{table}


\begin{prop}\label{abbacchio}
The probability that a randomly selected ranked species tree with $n$ internal nodes produces ARGTs approaches $1$ as $n \rightarrow \infty$.
\end{prop}
\emph{Proof.} Consider the number $c'_n + c''_n$ of ordered ranked trees of size $n$ that are MP at their root. Here $c'_n$ is the number of ordered ranked trees $t$ of size $n$ that are MP at their root and that have $H(t)=t$, and $c''_n$  is the number of ordered ranked trees of size $n$ that have $H(t) \neq t$ (and that are therefore MP at the root). The remaining $n! - c'_n-c''_n$ ordered ranked trees of size $n$ are NMP at the root. Observe that, if $n \geq 3$, then
\begin{equation}\label{pat}
c_{n+1}' =  \sum_{q=1}^{n-1} q! \,  (n-q) ! \, 2^{\min(q,n-q)} = n! \, \alpha_n.
\end{equation}
This result holds because each tree $t$ counted in $c'_{n+1}$ is built by appending two ordered ranked trees of sizes $1 \leq q \leq n-1$ and $n-q$ to a shared root. Once these subtrees are chosen, we choose one of the $2^{\min(q,n-q)}$ orderings that create an MP-subtree at the root of $t$ to merge the rankings of the subtrees of sizes $q$ and $n-q$. This value is obtained by noting from the definition of MP-subtrees that for $t$ to be MP at the root, the coalescence sequence for $t$ must have the form (\ref{potta}) once names $L$ and $R$ have been assigned to the two subtrees of the root in such a way that $|R| = \min(q,n-q)$. The number of sequences satisfying (\ref{potta}) is $2^{|R|} = 2^{\min(q,n-q)}$.

Moreover, we have
\begin{equation}\label{dente}
c''_{n+1} = 2 n!,
\end{equation}
because each tree counted in $c''_{n+1}$ has a leaf---a subtree of size $0$---appended to the root, and its other subtree of the root has size $n$. The factor of 2 arises because the leaf can appear on either side of the root.

By (\ref{collina}), because ranked species trees that are NMP at their root produce ARGTs, ${(n+1)! \, (n! -  c'_n - c''_n )}/{2^n}$ gives a lower bound for the number of ranked species trees of size $n$ producing ARGTs. Dividing by the number of ranked species trees of size $n$ (\ref{spectr}), by (\ref{pat}) and (\ref{dente}), we obtain
\begin{eqnarray}\nonumber
1 - \frac{c'_n + c''_n}{n!} &=& 1- \frac{(n-1)! \, \alpha_{n-1} + 2 (n-1)!}{n!} \\\nonumber
&=& 1 - \frac{\alpha_{n-1} + 2}{n}.
\end{eqnarray}
By (\ref{alf}), this value nears $1$ as $n$ becomes large.  $\,\,\, \Box$

\subsection{Ranked species trees $t$ that are NMP at the subtree $H(t)$} \label{acca}
We have shown that the fraction of ranked species trees $t$ that are NMP at subtree $H(t)$ approaches 1 as $n \rightarrow \infty$. In this section, we extend beyond this result to enumerate the set of ranked species trees that are NMP at $H(t)$. We achieve the result by counting ordered ranked trees $t$ that are MP at $H(t)$.

\begin{figure}
\begin{center}
\includegraphics*[width=.96\columnwidth]{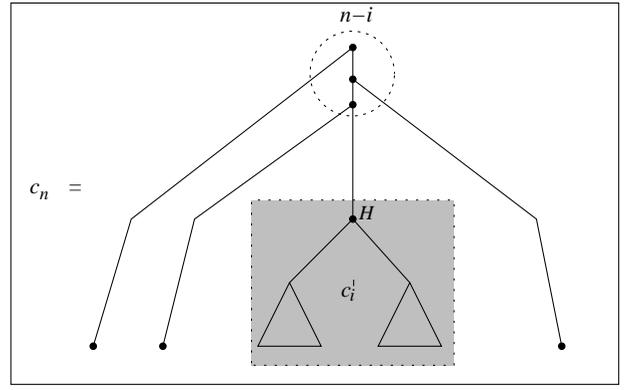}
\end{center}
\caption{{\small Decomposition of an ordered ranked tree $t$ of size $n$ that is in $\mathcal{R}_n^{(H)}$, is neither a caterpillar nor a pseudocaterpillar, and has subtree $H(t)$ maximally probable. The highlighted subtree is used for $c_i^{\prime}$ in computing (\ref{fero}).}}\label{castello}
\end{figure}

Let $c_n$ be the number of ordered ranked trees $t$ of size $n$ that are neither caterpillar nor pseudocaterpillar and that have the property that the subtree $H(t)$ is MP at its root. For $n \geq 4$, the smallest number of internal nodes for which a tree can be neither a caterpillar nor a pseudocaterpillar, we have
\begin{equation} \label{fero}
c_{n} =  \sum_{i=4}^{n} ({c}_{i}^{\prime})  2^{n-i},
\end{equation}
where $c_i^{\prime}$ is, as in the proof of Proposition~\ref{abbacchio}, the number of trees $t$ of size $i$ that are MP at their root and that have $H(t)=t$. The result is obtained by noting that each tree $t$ counted in $c_n$ is constructed from a tree in $c'_i$, with $4 \leq i \leq n$, which reaches the root of $t$ through a branch to which $n-i$ leaves are appended (Fig.~\ref{castello}). The leaves can be placed on either the right or the left of the branch, producing the factor $2^{n-i}$.

Observe that the number of caterpillar or pseudocaterpillar ordered ranked trees is given by
\begin{equation}\label{scatta}
|\mathcal{R}_n^{(cat)}| = 3 \cdot 2^{n-2}.
\end{equation}
In particular, we have $2^{n-1}$ caterpillar ordered ranked trees obtained by the possible left-right orientations of the leaves stemming from $n-1$ of the coalescences (all coalescences except the root of the cherry). Similarly, we have $2^{n-2}$ pseudocaterpillar ordered ranked trees, considering the two possible left-right orientations of all coalescences except the roots of the two cherries.  Therefore, $|\mathcal{R}_n^{(H)}| = c_n + |\mathcal{R}_n^{(cat)}| = c_n + 3 \cdot 2^{n-2}$.

The sequence $c_n^{\prime}$ can be computed as in (\ref{pat}). Using (\ref{fero}), we obtain
\begin{equation}\label{h}
c_n = \sum_{i=4}^{n}  (i-1)! \,  \alpha_{i-1} \,  2^{n-i} =  2^n \left( \sum_{i=4}^{n} (i-1)! \, \alpha_{i-1} \, 2^{-i} \right).
\end{equation}
Using $|\mathcal{R}_n^{(H)}|$ with (\ref{collina}), we can compute the number of ranked species trees $t$ that are MP at subtree $H(t)$.
\begin{prop}\label{lino}
The number of ranked species trees $t$ with $n$ internal nodes that are MP at the subtree $H(t)$ is
\begin{eqnarray}\label{mpH}
&& |\mathcal{S}_n^{(H)}| = \frac{(n+1)! \, |\mathcal{R}_n^{(H)}|}{2^n} \nonumber \\
&& = \frac{(n+1)! \, (c_n + 3 \cdot 2^{n-2} )}{2^n} \nonumber \\ 
&& = \frac{(n+1)!}{2^n} \nonumber \\
&& \quad \times \left[ 2^n \left( \sum_{i=4}^{n} (i-1)! \, \alpha_{i-1} \, 2^{-i} \right)  + 3 \cdot 2^{n-2} \right], \nonumber \\
\end{eqnarray}
where $\alpha_n$ can be computed as in (\ref{alfia}).

The number of ranked species trees $t$ with $n$ internal nodes that are NMP at the subtree $H(t)$ is
\begin{eqnarray}\label{nmpH}
&& |\mathcal{S}_n| - |\mathcal{S}_n^{(H)}|  = \frac{(n+1)!}{2^n} \nonumber \\
&& \quad \times \left[ n! - 2^n \left( \sum_{i=4}^{n} (i-1)! \, \alpha_{i-1} \, 2^{-i} \right) - 3 \cdot 2^{n-2}  \right]. \nonumber \\
\end{eqnarray}
\end{prop}

\noindent \textbf{Bounds.} By Proposition~\ref{lino}, the exact number of ranked species trees $t$ that are NMP at the subtree $H(t)$ can be computed. From Proposition~\ref{abbacchio}, the probability that a randomly selected ranked species tree $t$ is NMP at $H(t)$ approaches $1$ as $n$ grows large. Here we provide upper and lower bounds for the speed of convergence.

Observe that (\ref{fero}) implies that $c_n \geq c'_n$. Using (\ref{pat}) and (\ref{pinco}), we can write
\begin{eqnarray}\label{sopra}
&& \frac {|\mathcal{R}_n^{(H)}|}{n!} \geq \frac{c'_n}{n!} = \frac{(n-1)! \, \alpha_{n-1}}{n!} \nonumber \\
&& \geq \frac{(n-1)! \, 4/(n-1)}{n!}   = \frac{4}{n(n-1)} \geq \frac{4}{n^2}. 
\end{eqnarray}

On the other hand, given that $|\mathcal{R}_n^{(H)}| - c'_n$ counts a set of trees for which $H(t) \neq t$, we must have $|\mathcal{R}_n^{(H)}| - c'_n \leq c''_n = 2(n-1)!$, where $c''_n$ is as in (\ref{dente}) and corresponds to the number of ordered ranked trees $t$ with $H(t)\neq t$.
Dividing by $n!$ and using inequalities (\ref{pennarello}) and (\ref{pennarello1}) gives
\begin{eqnarray}\nonumber
\frac{|\mathcal{R}_n^{(H)}|}{n!} & \leq & \frac{c''_n + c'_n}{n!} = \frac{2(n-1)! + \alpha_{n-1} (n-1)!}{n!}  \\\nonumber
&=& \frac{2+\alpha_{n-1}}{n} \\ \nonumber
& \leq & \frac{1}{n} \left[ 2 + 2\left( \frac{\lfloor (n-1)/2 \rfloor + 10}{\lfloor (n-1)/2 \rfloor (\lfloor (n-1)/2 \rfloor -1)} \right)  \right] \\ \nonumber
& \leq &  \frac{1}{n} \left[ 2 + 2\left( \frac{ (n-1)/2  + 10}{ ((n-2)/2)  ( (n-2)/2  -1)} \right)  \right] \\
&=& \frac{2(n^2-4n+46)}{n(n-2)(n-4)}. \label{sotto}
\end{eqnarray}

When $n$ becomes large, the value
$$\frac{|\mathcal{S}_n \setminus \mathcal{S}_n^{(H)}|}{|\mathcal{S}_n|} = 1 -  \frac{|\mathcal{R}_n^{(H)}|}{n!},$$ that is, the probability that a randomly selected ranked species tree $t$ is NMP at $H(t)$, approaches $1$ at most as fast as $1-4/n^2$ (\ref{sopra}) and at least as fast as $1-2(n^2-4n+46)/[n(n-2)(n-4)]$ (\ref{sotto}).

\begin{figure}[tb]
\begin{center}
\includegraphics[width=.6\columnwidth,trim=0 0 0 0,angle=270]{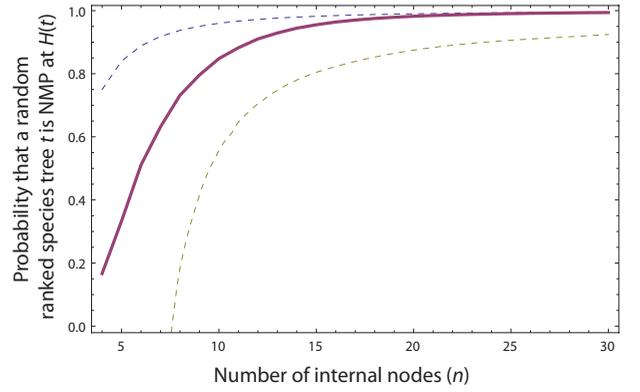}
\end{center}
\caption{{\small The probability that subtree $H(t)$ is non-maximally probable in a randomly selected ranked species tree $t$ with $n$ nodes, or $1-\mathcal{R}_n^{(H)}/{n!}$. The probability is confined by lower bound $1-2(n^2-4n+46)/[n(n-2)(n-4)]$ and upper bound $1-4/n^2$.}}\label{fast}
\end{figure}

Fig.~\ref{fast} plots the exact value of $1 -|\mathcal{R}_n^{(H)}|/n!$ with its bounds. The probability that a randomly selected ranked species tree $t$ is NMP at $H(t)$---and that it therefore produces ARGTs at $H(t)$---approaches 1 quickly.  Moreover, the upper bound appears to approximate the probability more accurately than does the lower bound.

\subsection{Ranked species trees that are NMP for at least one subtree}\label{double}

The set $\mathcal{S}_n \setminus \mathcal{S}_n^{(mp)}$---ranked species trees of size $n$ containing at least one NMP-subtree---is a superset of $\mathcal{S}_n \setminus \mathcal{S}_n^{(H)}$, and it thus expands the class of ARGT-producing ranked gene trees beyond the set $\mathcal{S}_n \setminus \mathcal{S}_n^{(H)}$. In this section we provide a recursion to compute the cardinality of $\mathcal{S}_n \setminus \mathcal{S}_n^{(mp)}$.
We also determine a more accurate lower bound for the number of ranked species trees that are ARGT-producing.

\begin{figure}
\begin{center}
\includegraphics*[scale=.86,trim=0 0 0 0]{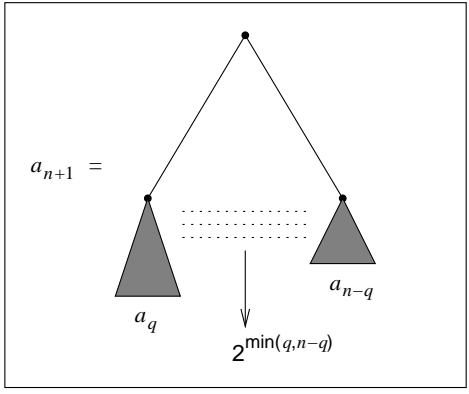}
\end{center}
\caption{{\small Decomposition of an ordered ranked tree of size $n+1$ that is in $\mathcal{R}_{n+1}^{(mp)}$ and has no non-maximally probable subtrees. The two subtrees of the root are taken from $\mathcal{R}^{(mp)}$, with sizes $q$ and $n-q$. According to the definition of maximally probable subtrees, once names $L$ and $R$ are assigned to the two subtrees in such a way that $|R| = \min(q,n-q)$, the number of possible rankings to obtain a sequence of coalescences satisfying (\ref{potta}) is $2^{|R|}$.}}
\label{recurs}
\end{figure}

We first focus on the class $\mathcal{R}_n^{(mp)}$ of ordered ranked trees of size $n$ avoiding NMP-subtrees.  Next, using (\ref{collina}), we convert the result to obtain $|\mathcal{S}_n \setminus \mathcal{S}_n^{(mp)}|$. Let $a_n = |\mathcal{R}_n^{(mp)}|$. Each tree in $\mathcal{R}_{n+1}^{(mp)}$ is obtained by appending to the same root two trees belonging to $\mathcal{R}^{(mp)}$, one of size $q$ and the other of size $n-q$, with $0 \leq q \leq n$. As was already noticed in the proof of Proposition~\ref{abbacchio}, when merging the rankings of the two subtrees of the root, exactly $2^{\min(q,n-q)}$ among the ${{n}\choose{q}}$ possible choices create an MP-subtree at the root. Recall that once we have assigned the names $L$ and $R$ to the two subtrees of the root in such a way that $|R| = \min(q,n-q)$, the number of possible rankings to obtain a sequence of coalescences of the form (\ref{potta}) is $2^{|R|}$. The decomposition is illustrated in Fig.~\ref{recurs}.

The recursion to compute $a_n$ is thus
\begin{equation}\label{pollo}
a_{n+1} =  \sum_{q = 0}^{n} (a_q a_{n-q}) 2^{\min(q,n-q)},
\end{equation}
where $a_0 = 1$. Taking $a_n$ and using property (\ref{collina}), we can obtain the cardinality of $\mathcal{S}_n^{(mp)}$.
\begin{prop}\label{labbro}
The number of ranked species trees with $n$ internal nodes that contain only MP-subtrees is
\begin{equation}\label{solomp}
|\mathcal{S}_n^{(mp)}| = \frac{(n+1)! \, |\mathcal{R}_n^{(mp)}| }{2^n}  = \frac{(n+1)! \, a_n }{2^n}.
\end{equation}

The number of ranked species trees with $n$ internal nodes that contain at least one NMP-subtree is
\begin{equation}\label{unnmp}
|\mathcal{S}_n| - |\mathcal{S}_n^{(mp)}| = \frac{(n+1)!}{2^n} \left( n! - a_n  \right).
\end{equation}
\end{prop}

An explicit formula for $|\mathcal{S}_n \setminus \mathcal{S}_n^{(mp)}|$ requires a solution of recursion (\ref{pollo}). Although we have not obtained such a solution, we can use the recursion to find a closed-form upper bound for $a_n$ and therefore a lower bound for the number of ranked species trees that produce ARGTs.  For large $n$, this bound is more accurate than the bound given by the number of ranked species trees $t$ that are NMP at subtree $H(t)$ (Proposition~\ref{lino}).

\medskip

\noindent \textbf{Bounds.} Fix a parameter $\beta$, $1/2 < \beta < 1$. Observe that
$$\sum_{q=0}^{\infty} \frac{q}{2^{q(2\beta - 1)}} = \sum_{q=1}^{\infty} \frac{q}{2^{q(2\beta - 1)}}$$
converges to a constant
\begin{equation}\label{serzio}
k_{\beta} = \frac{1}{2^{2\beta-1}\left(1-1/2^{2\beta - 1} \right)^2}.
\end{equation} 
This result is obtained by noting that $$\sum_{q=0}^{\infty} \frac{z^{q}}{v^q} = \frac{1}{1-z/v},$$ differentiating both sides with respect to $z$, setting $v = 2^{2\beta - 1}$, and choosing $z=1$. For instance, when $\beta = 6/11$, we have $k_{\beta} \approx 251.762$.

When $0 < q \leq n/2$, by (\ref{sturlo}), ${{n}\choose{q}} \geq {{2q}\choose{q}} \geq {2^{2q-1}}/{\sqrt{q}}$. For every $n$, we then have a bound:
\begin{eqnarray}\nonumber
\sum_{q=0}^{\lfloor n/2 \rfloor} \frac{2^q}{{{n}\choose{q}}^\beta} &=& 1+\sum_{q=1}^{\lfloor n/2 \rfloor} \frac{2^q}{{{n}\choose{q}}^\beta} \leq 1 + \sum_{q=1}^{\lfloor n/2 \rfloor} \frac{2^q}{{{2q}\choose{q}}^\beta} \\\nonumber
&\leq&  1 + \sum_{q=1}^{\infty} \frac{2^q}{{{2q}\choose{q}}^\beta}  \leq 1 +  \sum_{q=1}^{\infty} \frac{2^q}{ \left( \frac{2^{2q}}{2 \sqrt{q}} \right)^{\beta}} \\\label{partenza}
&=& 1 + 2^{\beta}\sum_{q=1}^{\infty} \frac{q^{\beta / 2}}{ 2^{q(2 \beta - 1)}} \leq 1+ 2^{\beta}k_{\beta}.
\end{eqnarray}

Choose $n$ to be a positive integer such that ${2(1+ 2^{\beta} k_{\beta})} \leq { (n + 1)^{\beta}}$. Let $c_{\beta} \geq 1$ be a constant such that for all $i$, $0 \leq i \leq n$, we have $a_i \leq c_{\beta}^i (i!)^{\beta}$. Note that the existence of $c_{\beta}$ is ensured because we could set, for instance, $c_{\beta} = \max\{a_i : 0 \leq i \leq n \}$. Thus, for such a constant we have both of the following conditions:
\begin{eqnarray} \label{condo2}
a_i & \leq & c_{\beta}^i (i!)^{\beta} \text{   for all $i$, $0 \leq i \leq n$} \\
\label{condo2b}
2(1+2^{\beta} k_{\beta}) & \leq & c_{\beta} (n+1)^{\beta}.
\end{eqnarray}

We can now prove by induction that if conditions (\ref{condo2}) and (\ref{condo2b}) are both satisfied for a certain $n$, then they also hold for $n+1$. For the second condition, the result is trivial. For the first condition, we use (\ref{pollo}):
\begin{eqnarray}\nonumber
a_{n+1} & \leq & 2 \sum_{q=0}^{\lfloor n/2 \rfloor} a_q a_{n-q} 2^q \leq 2 c_{\beta}^n \sum_{q=0}^{\lfloor n/2 \rfloor} (q!)^{\beta} (n-q)!^{\beta} 2^q  \nonumber \\
&=&  \frac{2 c_{\beta}^{n+1} (n+1)!^{\beta}}{c_{\beta}(n+1)^{\beta}} \sum_{q=0}^{\lfloor n/2 \rfloor} \frac{2^q}{{{n}\choose{q}}^\beta} \nonumber \\
&\leq & c_{\beta}^{n+1} (n+1)!^{\beta} \, \frac{2(1+2^{\beta} k_{\beta})}{c_{\beta} (n+1)^{\beta}} \nonumber \\
& \leq & c_{\beta}^{n+1} (n+1)!^{\beta}. \label{giollo}
\end{eqnarray}

We have therefore proven the following result.

\begin{prop}\label{chiarina}
Choose $\beta$ with $1/2 < \beta < 1$. Take a positive constant $c_{\beta}$ and define $$X=X(\beta,c_{\beta}) = \big[2(1+2^{\beta}k_{\beta})/c_{\beta} \big]^{1/\beta} - 1.$$ Suppose it can be verified that for every integer $n$ with $0 \leq n \leq X$,
\begin{equation}\label{dsa}
a_n \leq c_{\beta}^n (n!)^{\beta}.
\end{equation}

Then for every $n \geq 0$, $a_n\leq c_{\beta}^n (n!)^{\beta}$, and therefore, the number of ranked species trees with $n$ internal nodes that produce ARGTs is at least
\begin{equation}\label{Mcara}
|\mathcal{S}_n| - |\mathcal{S}_n^{(mp)}| \geq \frac{(n+1)!}{2^n}[n!-(c_{\beta})^n (n!)^{\beta}].
\end{equation}
\end{prop}

The upper bound for $a_n$ contained in Proposition~\ref{chiarina} shows that for $n$ large, the number of ranked species trees that contain only MP-subtrees is much smaller than the number of ranked species trees $t$ that are MP at $H(t)$. Indeed, from (\ref{sopra}) we have that $|\mathcal{R}_n^{(H)}| \geq {4}n!/{n^2}$, and therefore, for any $1/2 < \beta < 1$, $$\frac{|\mathcal{S}_n^{(H)}|}{|\mathcal{S}_n^{(mp)}|} \geq \frac{(4 \, n!)/n^2}{ c_{\beta}^n (n!)^{\beta}} = \frac{4(n!)^{1-\beta}}{n^2 c_{\beta}^n} \rightarrow \infty.$$
The constants $c_{\beta}$ in Proposition~\ref{chiarina} can be evaluated numerically. If we fix, for instance, $\beta=6/11$, then we have $k_{\beta} \approx 251.762$ as noted above. In this case, setting $c_{\beta}=c_{6/11}=5$, we have $X(\beta,c_{\beta}) \approx 9449.7$. We can then  computationally verify that condition (\ref{dsa}) is satisfied for every $n$, $0 \leq n \leq 9449$. Thus, with $\beta = 6/11$ and $c_{\beta} = 5$, (\ref{dsa}) holds for every $n\geq 0$. An efficient implementation of recursion (\ref{pollo}) can be achieved by saving each $a_j$ once computed, to minimize the number of calls to the recursive steps.

\section{Conclusions}

\begin{table*}[tb]
\vspace{-.2cm}
\caption{{\small Cardinalities of sets of ranked species trees with $n$ internal nodes.}}
\label{tavola1}
\fontsize{9}{11}\selectfont
\begin{center}
\begin{tabular}{|c| c | c c c c c c c | }\hline \label{tavola}
                                                &                 & \multicolumn{7}{c|}{$n$}                                          \\ \cline{3-9}
Class of trees                                  & Equation        & $4$ & $5$  & $6$   & $7$      & $8$        & $9$        & $10$         \\ \hline
$\mathcal{S}_n$                                 & (\ref{spectr})  & 180 & 2700 & 56700 & 1587600  & 57153600   & 2571912000 & 141455160000 \\
$\mathcal{S}_n \setminus \mathcal{S}_n^{(cat)}$ & (\ref{scatta2}) &  90 & 2160 & 52920 & 1557360  & 56881440   & 2569190400 & 141425222400 \\
$\mathcal{S}_n \setminus \mathcal{S}_n^{(mp)}$  & (\ref{unnmp})   &  30 & 900  & 30240 & 1083600  & 46176480   & 2278886400 & 132773256000 \\
$\mathcal{S}_n \setminus \mathcal{S}_n^{(H)}$   & (\ref{nmpH})    &  30 & 900  & 28980 & 1002960  & 41821920   & 2047096800 & 119964952800 \\
$\mathcal{S}_n^{(cat)}$                         & (\ref{scatta2}) &  90 & 540  & 3780  & 30240    & 272160     & 2721600    & 29937600     \\ \hline
\end{tabular}
\end{center}
\vspace{-.1cm}
{\small $\mathcal{S}_n^{(cat)}$ is the set containing caterpillar and pseudocaterpillar ranked species trees. $ \mathcal{S}_n \setminus \mathcal{S}_n^{(H)}$ is the set of trees $t$ for which the subtree $H(t)$ is NMP. $\mathcal{S}_n \setminus \mathcal{S}_n^{(mp)}$ is the set of trees containing at least one NMP-subtree. $\mathcal{S}_n \setminus \mathcal{S}_n^{(cat)}$ is the set of trees that are neither caterpillar nor pseudocaterpillar.  $\mathcal{S}_n$ is the set of ranked species trees.}
\end{table*}

We have examined three nested classes of ranked species trees (\ref{gerarchia}) characterized by the presence or absence of particular subtree patterns: $\mathcal{S}_n \setminus \mathcal{S}_n^{(H)}$, a class of ranked species trees proven by Degnan {\it et al.}~\cite{DegnanEtAl12:tcbb} to produce ARGTs; $\mathcal{S}_n \setminus \mathcal{S}_n^{(mp)}$, a larger class that by extension of their proof was identified as producing ARGTs; and the still larger class $\mathcal{S}_n \setminus \mathcal{S}_n^{(cat)}$ that excludes caterpillar and pseudocaterpillar ranked species trees proven by \cite{DegnanEtAl12:tcbb} \emph{not} to produce ARGTs.

Extending beyond the result of \cite{DegnanEtAl12:tcbb} that for each unranked species tree---with the exception of caterpillars and pseudocaterpillars---at least one ranking exists that gives rise to ARGTs, we have demonstrated that as $n \rightarrow \infty$, \emph{almost all} ranked species trees with $n$ internal nodes give rise to ARGTs (Proposition~\ref{abbacchio}). We have additionally provided a closed-form for the cardinality $|\mathcal{S}_n \setminus \mathcal{S}_n^{(H)}|$ (\ref{nmpH}) and a recursion as well as a closed-form lower bound for $| S_n \setminus \mathcal{S}_n^{(mp)} |$ (\ref{unnmp}, \ref{Mcara}).

For illustration, Table~\ref{tavola1} shows the cardinalities for small $n$, alongside the total number of ranked species trees $|S_n|$, the upper bound  $ |\mathcal{S}_n \setminus \mathcal{S}_n^{(cat)} |$ on the number of ranked species trees with ARGTs, and the lower bound $\mathcal{S}_n^{(cat)}$ on the number of ranked species trees without ARGTs.  The row for $\mathcal{S}_n \setminus \mathcal{S}_n^{(H)}$ extends a corresponding enumeration in Table 1 of \cite{DegnanEtAl12:tcbb}, correcting an error in the $n=7$ case ($n=8$ in \cite{DegnanEtAl12:tcbb}, which indexed cases by the number of leaves rather than the number of internal nodes).  It can be observed from the table that the quantities in the central row increase quite quickly with $n$ when considered as a fraction of $|\mathcal{S}_n|$.

\fil{The problem of characterizing the set of ranked species trees that produce ARGTs is analogous to the corresponding problem of characterizing the set of unranked species trees that produce anomalous unranked gene trees in the unranked case~\cite{DegnanAndRosenberg06, Rosenberg13:mbe, RosenbergAndTao08}. In that context, every species tree with four or more species, as well as the caterpillar species tree with four species, produces anomalous unranked gene trees \cite{DegnanAndRosenberg06}. Our work extends the analogy: for large $n$, not only does almost every unranked species tree have a ranking the produces anomalous ranked gene trees, almost every \emph{ranked} species trees produces anomalous ranked gene trees. The related characterization in the unranked case has been useful in facilitating the development of species tree inference methods and the design of simulation-based tests relying on unranked gene trees \cite{Rosenberg13:mbe}, and we expect our results to serve in a similar role in the ranked case.}
 
We note that we have not \fil{fully} completed the characterization of ranked species trees that produce ARGTs, a problem that was left open by \cite{DegnanEtAl12:tcbb}.  We have, however, shown that the work of \cite{DegnanEtAl12:tcbb} implies that among all ranked species trees with $n$ internal nodes, the fraction that produce ARGTs approaches 1---and approaches it quickly. Our recursion for $| \mathcal{S}_n \setminus \mathcal{S}_n^{(mp)} |$ as well as (\ref{nmpH}) and (\ref{Mcara}) provide lower bounds for the number of ranked species trees with $n$ internal nodes that are ARGT-producing. An upper bound is provided by the cardinality of the set of ranked species trees excluding only the caterpillars and pseudocaterpillars, or
\begin{equation} \label{scatta2}
|\mathcal{S}_n \setminus \mathcal{S}_n^{(cat)} | = [(n+1)!/2^n](n! - 3 \cdot 2^{n-2}),
\end{equation}
where $|\mathcal{S}_n^{(cat)} | = [(n+1)!/2^n] ( 3 \cdot 2^{n-2})$. For the unsolved complete characterization of ranked species trees that produce ARGTs, the exact value must lie in a narrow range bounded between $| \mathcal{S}_n \setminus \mathcal{S}_n^{(mp)} |$ and $ |\mathcal{S}_n \setminus \mathcal{S}_n^{(cat)} |$.


\ifCLASSOPTIONcompsoc
  \section*{Acknowledgments}
\else
  \section*{Acknowledgment}
\fi
 We acknowledge grant support from the National Science Foundation (DBI-1146722) and the Burroughs Wellcome Fund.

\ifCLASSOPTIONcaptionsoff
  \newpage
\fi

\IEEEtriggercmd{\enlargethispage{-5in}}

\vfill



\begin{thebibliography}{1}

\bibitem{AllmanDegnanRhodes11:jtb} 
E.S.~Allman, J.H.~Degnan, J.A.~Rhodes.
Determining species tree topologies from clade probabilities under the coalescent.
\emph{Journal of Theoretical Biology} 289: 96-106.

\bibitem{AllmanDegnanRhodes11:jmb} 
E.S.~Allman, J.H.~Degnan, J.A.~Rhodes.
Identifying the rooted species tree from the distribution of unrooted gene trees under the coalescent.
\emph{Journal of Mathematical Biology} 62: 833-862.

\bibitem{Brown94} 
J.K.M.~Brown.
Probabilities of evolutionary trees.
\emph{Systematic Biology} 43: 78-91.

\bibitem{Bullen98} 
P.S.~Bullen.
A Dictionary of Inequalities.
Harlow, UK: Addison Wesley Longman (1998).

\bibitem{Degnan13} 
J.H.~Degnan.
Anomalous unrooted gene trees.
\emph{Systematic Biology} 62: 574-590 (2013).

\bibitem{DegnanEtAl09} 
J.H.~Degnan, M.~DeGiorgio, D.~Bryant, N.A.~Rosenberg.
Properties of consensus methods for inferring species trees from gene trees. 
\emph{Systematic Biology} 58: 35-54 (2009).

\bibitem{DegnanAndRosenberg06} 
J.H.~Degnan, N.A.~Rosenberg.
Discordance of species trees with their most likely gene trees.
\emph{PLoS Genetics} 2: 762-768 (2006).

\bibitem{DegnanAndRosenberg09} 
J.H.~Degnan, N.A.~Rosenberg.
Gene tree discordance, phylogenetic inference and the multispecies coalescent.
\emph{Trends in Ecology and Evolution} 24: 332-340 (2009).

\bibitem{DegnanEtAl12:tcbb} 
J.H.~Degnan, N.A.~Rosenberg, T.~Stadler.
A characterization of the set of species trees that produce anomalous ranked gene trees.
\emph{IEEE/ACM Transactions on Computational Biology and Bioinformatics} 9: 1558-1568 (2012).

\bibitem{DegnanEtAl12:mathbiosci} 
J.H.~Degnan, N.A.~Rosenberg, T.~Stadler.
The probability distribution of ranked gene trees on a species tree.
\emph{Mathematical Biosciences} 235: 45-55 (2012).

\bibitem{DegnanAndSalter05} 
J.H.~Degnan, L.A.~Salter.
Gene tree distributions under the coalescent process.
\emph{Evolution} 59: 24-37 (2005).

\bibitem{DisantoEtAl13}
F.~Disanto, A.~Schlizio, T.~Wiehe.
Yule-generated trees constrained by node imbalance.
\emph{Mathematical Biosciences} 246: 139-147 (2013).

\bibitem{Edwards70} 
A.W.F.~Edwards.
Estimation of the branch points of a branching diffusion process.
\emph{Journal of the Royal Statistical Society Series B} 32: 155-174 (1970).

\bibitem{FlajoletAndSedgwick09} 
P.~Flajolet, R.~Sedgewick.
Analytic Combinatorics.
Cambridge: Cambridge University Press (2009).

\bibitem{Harding71} 
E.F.~Harding.
The probabilities of rooted tree-shapes generated by random bifurcation.
\emph{Advances in Applied Probability} 3: 44-77 (1971).

\bibitem{HeinEtAl05} 
J.~Hein, M.H.~Schierup, C.~Wiuf. 
Gene Genealogies, Variation and Evolution.
Oxford: Oxford University Press (2005).

\bibitem{Maddison97} 
W.P.~Maddison.
Gene trees in species trees.
\emph{Systematic Biology} 46: 523-536 (1997).

\bibitem{Page91} 
R.D.M.~Page.
Random dendograms and null hypotheses in cladistic biogeography.
\emph{Systematic Zoology} 40: 54-62 (1991).

\bibitem{PamiloAndNei88} 
P.~Pamilo, M.~Nei.
Relationships between gene trees and species trees.
\emph{Molecular Biology and Evolution} 5: 568-583 (1988).

\bibitem{PetkovsekEtAl96} 
M.~Petkov\v{s}ek, H.S.~Wilf, D.~Zeilberger.
A=B.
Wellesley, MA: Peters (1996).

\bibitem{Rosenberg02} 
N.A.~Rosenberg.
The probability of topological concordance of gene trees and species trees.
\emph{Theoretical Population Biology} 61: 225-247 (2002).

\bibitem{Rosenberg06:anncomb} 
N.A.~Rosenberg.
The mean and variance of the numbers of $r$-pronged nodes and $r$-caterpillars in Yule-generated genealogical trees.
\emph{Annals of Combinatorics} 10: 129-146 (2006).

\bibitem{Rosenberg13:mbe} 
N.A.~Rosenberg.
Discordance of species trees with their most likely gene trees: a unifying principle.
\emph{Molecular Biology and Evolution} 30: 2709-2713 (2013).

\bibitem{RosenbergAndTao08} 
N.A.~Rosenberg, R.~Tao.
Discordance of species trees with their most likely gene trees: the case of five taxa. 
\emph{Systematic Biology} 57: 131-140 (2008).

\bibitem{Song06} 
Y.S.~Song.
Properties of subtree-prune-and-regraft operations on totally-ordered phylogenetic trees.
\emph{Annals of Combinatorics} 10: 147-163 (2006).

\bibitem{StadlerAndDegnan12} 
T.~Stadler, J.H.~Degnan.
A polynomial time algorithm for calculating the probability of a ranked gene tree given a species tree.
\emph{Algorithms for Molecular Biology} 7: 7 (2012).


\bibitem{SteelAndMcKenzie01} 
M.~Steel, A.~McKenzie.
Properties of phylogenetic trees generated by Yule-type speciation models.
\emph{Mathematical Biosciences} 170: 91-112 (2001).

\bibitem{Tajima83} 
F.~Tajima.
Evolutionary relationship of DNA sequences in finite populations.
\emph{Genetics} 105: 437-460 (1983).

\bibitem{Wakeley09} 
J.~Wakeley. 
Coalescent Theory: An Introduction.
Greenwood Village, CO: Roberts (2009).

\bibitem{Wu12}
Y.~Wu. 
Coalescent-based species tree inference from gene tree topologies under incomplete lineage sorting by maximum likelihood.
\emph{Evolution} 66: 763-775 (2012).

\bibitem{Yule24}
G.U.~Yule.
A mathematical theory of evolution based on the conclusions of Dr.~J.~C.~Willis, F.~R.~S.
\emph{Philosophical Transactions of the Royal Society of London Series B} 213: 21-87 (1924).






\end{thebibliography}
\end{document}